\documentstyle[aps,prl,twocolumn]{revtex}
\input epsf 
\begin{document}
\title{Selection of the Saffman-Taylor Finger Width\\
 in the Absence of Surface Tension: an Exact Result    }
  \author {Mark Mineev-Weinstein }
  \address{Theoretical Division, MS-B213, LANL,
        Los Alamos, NM 87545, USA }
\maketitle
\begin{abstract}
Using exact time-dependent non-singular solutions [Mineev-Weinstein and Dawson, Phys.~Rev. {\bf E 50}, R24 (1994);  Dawson and Mineev-Weinstein, Physica {\bf D 73}, 373 (1994)], we solve the Saffman-Taylor finger selection problem in the absence of surface tension. We show that a generic interface in a Hele-Shaw 
cell evolves to a non-linearly stable single uniformly advancing finger occupying one half of the channel width. This result contradicts the generally accepted belief that surface tension is indispensable for the selection of the $\frac{1}{2}$-width finger.
\end{abstract}
\pacs{PACS numbers 47.15.Hg, 47.20.Hw, 68.10.-m, 68.70.+w }
The problem of the finger width selection was posed in 1958 by Saffman 
and Taylor \cite{1} in their study of displacement of oil by water in a 
Hele-Shaw cell. This cell consists of two parallel glass plates separated 
by a thin gap occupied by a viscous liquid which is pushed by a less 
viscous one. This simple device is useful for modeling flows in porous 
media, the study of which is vitally important for many applications.
Flows in uniform porous media and in the Hele-Shaw
cell  are described by the same Darcy law:
$ \mbox{\bf v} = - \nabla p $,
where {\bf v} is fluid velocity, and $p$ is pressure.
For the Hele-Shaw cell, this equation follows from the Stokes equation 
averaged over the direction perpendicular to the parallel plates.

Saffman and Taylor \cite{1} observed that an almost planar initial oil/water 
interface becomes unstable and gives rise to many competing fingers which 
eventually evolve to a single uniformly advancing finger occupying {\it one 
half} of the channel width, if the surface tension is very small.  But, as 
was analytically shown in the same paper \cite{1}, {\it any finger width} 
is possible. So the selection problem was stated: why does Nature choose 
the width of one-half?

This problem appeared to be universal, i.e.\ the same selection phenomenon is
common for displacement of various viscous liquids by less viscous ones for 
immiscible incompressible liquids. This problem is related to the problem of 
pattern selection in nonequilibrium phenomena, which has been of much 
subsequent interest \cite{3}.

When the viscosity of water is negligible compared with oil viscosity, the 
mathematical formulation of this problem in the absence of surface tension 
has the form:
\begin{eqnarray}
\left\{
\begin{array}{l} 
\nabla^2 p = 0\quad {\rm (in~the\,oil~domain),}\\ 
p =  - x\quad {\rm if}~~x\,\to +\infty \, \quad {\rm 
(oil~pushed~to~the~right),}\\ 
p = 0 \quad {\rm(at~ the~ oil/water~interface),} 
\label{eq2}
\\ 
\partial_n p = 0 \quad {\rm for}~~y = \pm\pi \,\quad 
\rm{(at~the~channel~walls),}\\ 
V_n =- \partial_n p \quad \rm{(at~the~oil/water~interface), } 
\end{array}
\right.
\end{eqnarray}
where $V_n$ is the normal velocity of the  interface,
$\partial_n p$ is the normal component of the pressure gradient, and
the channel width is chosen to be 2$\pi$ in our scaled units.

The solution of the system (\ref{eq2}), describing a finger moving in the 
$x$-direction with velocity $1/\lambda$ and occupying the portion $\lambda$ 
of the Hele-Shaw channel width, is \cite{1} 
\begin{equation}\label{eq3}
x = 2(1-\lambda) \log [\cos({y}/{2\lambda})] + t/\lambda \ .
\end{equation}

We parametrize the moving interface $z(t, \phi) = x(t, \phi) + iy(t, \phi)$ 
at time $t$ by the  parameter $\phi \, \epsilon \, [0, 2\pi]$. 
After the shift $y \rightarrow y - \pi\lambda$, 
Eq.(2)
can be rewritten as
\begin{equation}\label{eq4}
 z(t, \phi) = {t}/{\lambda} + ( 1-\alpha)i\phi + \alpha
\log [ ({e^{i\phi} - 1})/{2i}]\,,
\end{equation}
where $\lambda=1-\alpha/2$.  The system (1) can be reduced to what we call 
the Laplacian growth equation (LGE) for the moving front $z(t, \phi)$ (see 
\cite{ms_sm} and references therein): 
\begin{equation}
\label{eq5}
{\rm Im}(\bar z_t z_{\phi}) = 1\,\,.
\end{equation}
Here the bar denotes complex conjugate, $z_t$ and $z_{\phi}$ are partial
derivatives, and the map $z(t, \phi)$ is conformal for Im $\phi \leq 0$.
One can easily see that $z(t, \phi)$ given by (3) is the traveling-wave 
solution of the LGE given by (\ref{eq5}). The finger width $\lambda$ is 
here a free parameter, while experimentally it is always $\frac{1}{2}$. 
What determines $\lambda$?

In \cite{1} Saffman and Taylor proposed that surface tension between the 
two fluids would solve the selection problem. Since then, it has been widely 
accepted that {\it the inclusion of surface tension is the only way to select 
the most stable finger width}, and much work was done toward solving the 
selection problem in this way (see books \cite{pelce} and references 
therein). While mathematically non-trivial and challenging \cite{pelce}, 
this activity, especially intensive in the 1980s, was nonetheless 
successfully completed \cite{pelce} and summarized in \cite{kkl}. In short, several groups in 1986-1987 \cite{2} confirmed (using expansion ``beyond all orders'' and reduction to a nonlinear eigenvalue problem) numerical evidence \cite{vdb} of the discrete spectrum of $\lambda$, decreasing to 1/2, in the limit  of low surface tension. Surface tension was claimed to be responsible 
for the selection: equating it to zero would make this analysis senseless.

We see two reasons for the absence of attempts to explain the selection 
without the inclusion of surface tension.  
First, because of the absence of analytic time-dependent solutions, all 
selection studies were focused on {\it linear stability} analysis of a 
{\it steady-state} traveling finger in the presence of surface tension, 
which is the main physical factor neglected when the continuous family 
of fingers (2) was derived \cite{1}. The second reason to include surface 
tension stems from the observation that almost all exact zero surface tension 
solutions of this problem obtained before 1994 \cite{bs84} exhibit finite-time 
singularities (cusps). Due to the belief that these solutions are general and 
thus capture main features of this problem, it was concluded that to reach 
long times, it is necessary to include surface tension to eliminate 
singularities \cite{bklst86}.  

In 1994, we reported \cite{ms_sm} a new class of exact time-dependent 
solutions of the LGE (\ref{eq5})  having the form:
\begin{equation}\nonumber 
z(t, \phi) = \tau(t) +i\mu \phi + \sum_{k=1}^{N}\alpha_k 
\log(e^{i\phi} - a_k(t)) \,,
\end{equation}
where $ \mu = 1-\sum_{k=1}^{N}\alpha_k$, $\alpha_k = {\rm const}$,
and $|a_k| < 1$.
With some (quite modest) constraints on $\{\alpha_k\}$, these
solutions remain non-singular and analytic for all times (no cusps) 
\cite{ms_sm}. The time dependence of $a_k(t)$ and $\tau(t)$ is given by
\begin{eqnarray}
&&\beta_k = z(t, i\,\log \bar a_k) = 
\tau - (1-\sum_{l=1}^{N}\alpha_l)\log \bar a_k \\
&&+\sum_{l=1}^{N}\alpha_l\log(\frac{1}{\bar a_k} - a_l) =
\rm{const, \,\,\,and}\,\nonumber  \\ 
&&t + C = \Big(1-\frac{1}{2} \sum_{k=1}^N \alpha_k \Big) \tau   +
\frac{1}{2}\sum_{k=1}^{N} \alpha_k  \log(a_k) \,,
\end{eqnarray}
where $k = 1, 2, ..., N$ and $C$ is a constant in time \cite{foot1}. 
Eqs. (6) and (7) follow from the substitution of (5) into (4).

All $a_k$ are located inside the unit circle and, if the same holds for 
the roots of $z_\phi$, then $z(t,\phi)$ is conformal for Im~$\phi~\leq 0$. 
We called these solutions $N$-finger 
solutions, since they 
describe the evolution of $N$ fingers. The class of solutions
(5) contains all previously known exact solutions \cite{bs84},\cite{s59},
\cite{h86}, including those which diverge in finite time. The subclass of 
these solutions 
without finite-time singularities, was also shown to be dense in the space of 
all analytic curves \cite{mmr}. {\it The dynamics of an arbitrary initial 
interface can be faithfully described within this class}. In addition to 
possessing these attractive mathematical properties, these solutions are very 
physical: they describe tip-splitting, side-branching, competition, 
coarsening and screening of growing fingers which are observed in all known 
experiments and simulations. 

\begin{figure}

\qquad \epsffile{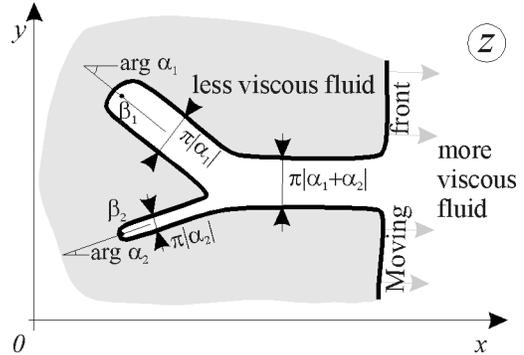} 

\smallskip
\caption{\label{grooves}Geometrical interpretation of the complex 
constants of motion $\alpha_k$ and $\beta_k$; $k=1, ..., N$.}
\end{figure}

The following geometrical interpretation of the constants $\{\alpha_k\}$ and 
$\{\beta_k\}$  \cite{ms_sm} is of great help: the complex number $\beta_k - 
\alpha_k \log 2$ is the location of the $k$-th stagnation point which the 
interface does not cross (tips of white grooves in Fig.~\ref{grooves} and
Fig.~\ref{4pi}), but 
approaches exponentially slowly, namely proportional to $\exp(-\tau/{\rm Re } 
\,\alpha_k)$. Near the $k$-th stagnation point, a groove with parallel walls 
originates, with width $\pi|\alpha_k|$ and angle with respect to the horizontal
axis $\theta_k=\arg\,\alpha_k$. In terms of these stagnation points and grooves
given by constants $\{\beta_k\}$ and $\{\alpha_k\}$, all the dynamical features
mentioned above are especially clear (see Fig.~\ref{grooves}). These grooves 
merge and finally  coalesce to a single growing finger in accordance with all 
known experiments and simulations \cite{1,pelce}. Formally this means that a 
generic initial interface given by the $N$-finger solution (5) necessarily 
evolves to a single uniformly advancing finger (see \cite{mkp} for details).

In this paper, we will solve the selection problem analytically, starting from 
first-principles, in the absence of surface tension ({\sl while interfacial 
tension is required to linearly stabilize the finger, as previously shown} 
\cite{pelce,kkl,2}). 
It is known \cite{s59} that the development of a single finger with a relative 
width $\lambda = 1 - \alpha/2$ in the long-time limit is described by
\begin{equation}
z(t,\phi) = \tau(t) + \mu i\phi +
\alpha \log\left(e^{i\phi}-a(t)\right)\;,
\label{new1}
\end{equation}
where $0<\alpha<2$, $0<a<1$. For $t\rightarrow\infty\;$, $\;\tau = 
2t/(2-\alpha)$ and $a(t) = 1-c \exp[-2t/\alpha(2-\alpha)]\,$, where $c$ and 
$\alpha$ are constants in time. Choosing the width of the Hele-Shaw cell to 
be $2\pi$, we have $z(t,2\pi) - z(t,0) = 2\pi i$, because of the periodic 
boundary conditions. Calculating $z(t,2\pi)-z(t,0)$ from (\ref{new1}) and 
using the fact that $|a|<1$, we obtain that $2\pi i = 2\pi i\mu + 2\pi 
i\alpha = 2\pi i(\mu+\alpha)$, or finally 
\begin{equation}
\mu + \alpha = 1\ .
\label{new2}
\end{equation}
Then we note that the second term in the right-hand side of (\ref{new1}), 
namely $\mu i\phi$, is the limiting value of the logarithm with a logarithmic
pole $\epsilon$ located at zero:
\begin{equation}
\mu i\phi = \mu \lim_{\epsilon \to 0}
\log\left(e^{i\phi}-\epsilon\right)\;.
\label{new3}
\end{equation}
Let us perturb the interface (8), corresponding to $\epsilon=0$, by the 
initially small non-zero $\epsilon$, $0 < \epsilon \ll 1$. The perturbed 
interface will have the form
\begin{equation}
z(t,\phi) = \tau + \mu \log\left(e^{i\phi}-\epsilon\right) +
\alpha \log\left(e^{i\phi}-a\right)\;,
\label{new4}
\end{equation}
Now one can easily see that the value $\epsilon = 0$ (and thus the 
finger described by (8) with $\mu \neq 0$) is unstable.  
The point is that Eq. (\ref{new4}) is exactly the $N = 2$ case of the 
$N$-finger solution (5) of the LGE (4).  
As one can see from (6) and (7), $a(t)$ and $\epsilon(t)$ merge at unity 
when $t\rightarrow\infty$:
\begin{equation}\label{new6}
a = 1 - l_1 e^{-\tau}\,, \quad
\epsilon = 1 - l_2 e^{-\tau}\,, 
\end{equation}
where the constants $l_1$ and $l_2$ are determined by
\begin{equation}\left\{\begin{array}{ccc}
\beta_1 &=& \mu \log(l_1 + l_2) + \alpha \log(2 l_1)\;,
\label{new7}\\
\beta_2 &=& \mu \log(2 l_2) + \alpha \log(l_1 + l_2)\;.
\end{array}\right.
\end{equation}
In view of (\ref{new6}), we substitute $1$ for both $a$ and $\epsilon$
with the accuracy $O(e^{-\tau})$, and thus obtain
from (\ref{new4}), for $t\rightarrow\infty$
\begin{equation}
z(t,\phi) = \tau + (\mu + \alpha)\log\left(e^{i\phi} - 1\right)\;.
\label{new8}
\end{equation}
Let us interpret the result (\ref{new8}): due to
the instability of the initial finger (\ref{new1}) with
width $\lambda = 1 - \alpha/2$, the new finger described by (\ref{new8})
has been formed. Its width is
\begin{equation}
\lambda^{\rm new} = 1 - ({\mu + \alpha})/{2} = {1}/{2}\,,
\label{new9}
\end{equation}
in accordance with the condition (\ref{new2}) that $\mu + \alpha = 1$.

Let us perturb the finger (\ref{new1}) in a more general
way, than we did in (11). We note that 
$$\mu i \phi = \lim_
{{\rm all~ }\epsilon_k\to 0}\;\sum_{k=1}^{N} \delta_k \log\left(e^{i\phi} - 
\epsilon_k\right)\,
$$ 
if $\sum_{k=1}^{N}\delta_k = \mu\;$.
Choosing all $\epsilon_k$ to be nonzero, we rewrite the finger (\ref{new1})
in a perturbed way as
\FL
\begin{equation}
z(t,\phi) = \tau + \sum_{k=1}^{N}\delta_k
\log\left(e^{i\phi} - \epsilon_k\right)
+ \alpha\log\left(e^{i\phi} - a\right).
\label{new11}
\end{equation}
Equation (\ref{new11}) is the $(N+1)$-finger solution (5) of the LGE (4) 
with dynamical conditions (6) and (7). Because of the density of the subclass 
of smooth solutions given by (5) we conclude that (\ref{new11}) describes a 
general perturbation of the finger (\ref{new1}), if $N$ is large enough.
As follows from (6) and (7), generally all logarithmic poles in the absence
of finite-time singularities {\it merge} [similarly to (12) for $N = 2$] in 
the long time limit to $1$ with exponential accuracy $O(e^{-\tau})$ \cite{mkp}, where
\begin{equation}
t= \tau\Big(1 - \frac{1}{2}\Big(\sum_{k=1}^{N}\delta_k + \alpha\Big)\Big)\;.
\label{new12}
\end{equation}
Because of this merging near the unit circle we substitute 1 for all 
$\epsilon_k(t)$ and $a(t)$ in (\ref{new11}) when $t \to \infty$ and obtain
\begin{equation}
z(t, \phi) = \tau + \Big(\sum_{k=1}^{N}\delta_k + \alpha\Big) 
\log\left(e^{i\phi} - 1\right)\;.
\label{new13}
\end{equation}
This formula describes the single finger formed from (8) under the 
perturbation (\ref{new11}). Its width is
\begin{equation}
\lambda^{\rm new} = 1 - \Big(\sum_{k=1}^{N}\delta_k + \alpha\Big)/{2}\,,
\label{new14}
\end{equation}
which is exactly one half since $\sum_{k=1}^{N}\delta_k = \mu$ (see above) 
and $\mu + \alpha = 1$ by an argument analogous to (\ref{new2}), so that
\begin{equation}
\lambda^{\rm new} = 1 - (\mu + \alpha)/2 = {1}/{2}\ .
\label{new15}
\end{equation}
Both (\ref{new15}) and ({\ref{new9}) indicate that, for obtaining instability 
of non-$\frac{1}{2}$-width finger and formation of the $\frac{1}{2}$-width 
finger in a long-time limit, surface tension is not needed. 

Now we start from an arbitrary initial interface in terms of non-singular 
solutions (5), (Re $\alpha_k > 0$). Because of the coalescence of all 
initially non-zero poles $a_k$ at the unit circle, the long-time limit of (5) 
is given by a finger (3) with a width of $(2-\sum_{k=1}^{N}\alpha_k)/2$. 
The $N$-finger solution (5) is the limit of $(N+k)$-finger solution expressed 
by the same equation, but without the term $\mu i \phi$. This limit corresponds
to equating $k$ of the poles to zero, and this value of zero can be easily 
shown (in the same way as above) to be unstable for all of these $k$ poles.
Because of the density of these solutions in the class of all analytic curves 
\cite{mmr}, this $(N+k)$-finger solution can be arbitrarily close to any 
analytic interface. Again, all $a_k(t)$ merge to 1 in the limit $t \to \infty$ 
[as stated earlier and proven in \cite{mkp}]. Thus we have in the long-time limit 
$$
z(t,\phi) = \tau + (\sum_{k=1}^{N+k}\alpha_k)\log\left(e^{i\phi} - 1\right) =
\tau + \log\left(e^{i\phi} - 1\right)\ .
$$
Here we used $\sum_{k=1}^{N+k}\alpha_k = 1$, since $2\pi i = z(t, 2\pi) - 
z(t, 0)$. So, we have demonstrated that initial interfaces evolving to the 
non-$\frac{1}{2}$-width finger are unstable with respect to formation of the 
$\frac{1}{2}$-width finger, which thus is shown to be the only attractor for 
all generic moving fronts in the Hele-Shaw cell represented by (5). (Solutions 
with several parallel fingers forming in asymptotics can also be easily shown 
to be unstable). The dynamics of the transition from an arbitrary interface to 
the $\case{1}{2}$-width finger is exactly described by the set of transcendental
equations (6) and (7) which involve only elementary functions. This selection 
of one half is  in  agreement with known experiments and simulations in the 
limit of low surface tension.

Now we will extend these results obtained for periodic boundary conditions
to the more physical ``no-flux'' boundary conditions (no flow across the 
lateral boundaries of the channel). This requires that the moving interface 
orthogonally intersects the walls of the Hele-Shaw cell. However, unlike the 
case of periodic boundary conditions, the end points at the two boundaries do 
not necessarily have the same horizontal coordinate. This is also a periodic 
problem where the period equals twice the width of the Hele-Shaw cell. The 
analysis is the same as before, but now only half of the strip should be 
considered as the physical Hele-Shaw channel, while the second half is the 
unphysical mirror image (see Fig.~\ref{4pi}).

To be brief we will perform the analysis for $N = 2$, but one can trivially 
extend it to an arbitrary $N$. The easily obtainable extension of (\ref{new1}) 
for the development of a single finger with the width $\lambda = 1 - (\alpha_1 
+ \alpha_2)/2$ is 
\begin{eqnarray}\nonumber 
&&z(t,\phi) = \tau(t) + \mu i\phi +
\alpha_1 \log\left(e^{i\phi}-a_1(t)\right) \\
&& + \alpha_2 \log\left(e^{i\phi}+a_2(t)\right)\;,
\label{new19}
\end{eqnarray}
where $0 < a_{1,2}(t) < 1$. 
The generalization of (9) is 
\begin{equation}
\mu = 1 -  (\alpha_1 + \alpha_2)
\label{new20} .
\end{equation}
\begin{figure}
\qquad \epsffile{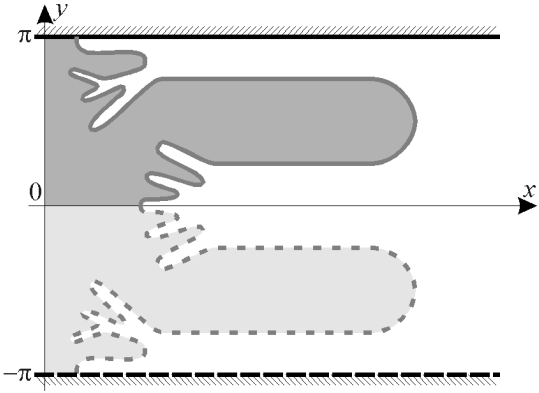}
\smallskip
\caption{\label{4pi} Only half of the periodic cell
should be considered as the physical Hele-Shaw channel, while 
the second half is the unphysical mirror image.}
\end{figure}
We note that 
\begin{eqnarray}
&&\mu i\phi = (\frac{1}{2} - \alpha_1 +\delta) \lim_{\epsilon_1\to
  0+}\log\left(e^{i\phi}-\epsilon_1\right) \\ \nonumber
&&+ (\frac{1}{2} - \alpha_2 -\delta) \lim_{\epsilon_2\to 0+}\log\left(e^{i\phi} + \epsilon_2\right)\;.
\label{new21}
\end{eqnarray}
Substituting this into (\ref{new19}) and allowing $\epsilon_{1,2}$
to be functions of time, we note that $z(t,\phi)$ is the $N =4$ solution (5)
of Eq.(4), where $\delta$ and $\epsilon_{1,2}$ must be real. 
If $|\delta| < 1/2$, all four logarithmic singularities $a_{1,2}(t)$ and $\epsilon_{1,2}(t)$ merge to 1. 
So we have in the long-time limit
\begin{eqnarray}\label{new24}
&&z(t,\phi) = \tau(t) + (\frac{1}{2} + \delta)\log\left(e^{i\phi}-1\right) \\
\nonumber 
&&+ (\frac{1}{2} - \delta) \log\left(e^{i\phi} + 1\right)\;.
\end{eqnarray}
This describes a finger moving between two grooves with widths $\pi(1 + 2\delta)/2$ and $\pi(1 - 2\delta)/2$ 
respectively (see the geometrical interpretation above). Thus the portion of the channel width occupied by the moving finger is 
\begin{equation}
\lambda^{\rm new} = \frac{2\pi - \pi(1 + 2\delta)/2 - \pi(1 - 2\delta)/2}{2\pi}  = \frac{1}{2}\,,
\label{new25}
\end{equation}
as before. So, for the no-flux boundary conditions, we have obtained the $\frac{1}{2}$-width
finger as expected. In experiments, the finger in the long-time limit is 
centralized in sense that axes of symmetry  of the finger 
and Hele-Shaw cell coincide. This corresponds to the condition $\delta = 0$. 

The finger with a width of one half is {\bf nonlinearly} {\sl
stable with respect to generic perturbation of logarithmic type (5).}
Namely, the shape of the finger can be destroyed at the initial (linear) 
stage, but eventually the $\frac{1}{2}$-width finger will be restored, 
because of the coalescence described above. (Of course, the $\frac{1}{2}$-width
finger is {\bf linearly} {\sl unstable without interfacial tension in accordance
with previous studies} 
\cite{pelce,kkl,2}.)

Regarding surface tension, we think that while mathematically still singular 
(because a small number multiplies the highest derivative), physically surface 
tension is a {\it regular } perturbation for this problem (unless very high 
curvatures exist which surface tension suppresses). 

In conclusion, we have analytically solved the finger selection problem in the 
absence of surface tension. By using the non-singular exact solutions of the 
LGE (4), we have demonstrated that the $\frac{1}{2}$-width finger is the only 
attractor for all generic moving fronts in a Hele-Shaw cell.

The author thanks H. Makaruk, W. MacEvoy, J. Pearson, V. Lvov, I. Procaccia,
M. Feigenbaum, D. H. Sharp for helpful discussions.

{\it Note added after publication:}
In the recent paper by A. P. Aldushin and B. J. Matkowsky 
(to be published in Appl. Math. Lett.) the selection of 
$\lambda = 1/2$ in the absence of surface tension was shown
to be consistent with minimum of assumed functional for the
steady-state solution.


\begin{references}
\bibitem{foot1} Particular examples of this class, limited however to a
specific symmetry, were known earlier \cite{s59}--\cite{h86}. The
solution (5), also has the symmetry such that the sum in the RHS of (5)
contains both terms with $\alpha_k$ and $a_k$ and with $\bar \alpha_k$
and $\bar a_k$ (unless both $\alpha_k$ and $a_k$ are real). It is
related to no-flux boundary conditions (see discussion and analysis in 
this paper beginning at "Now we will extend these results.." and in
Ref. \cite{mkp}).
\end{references}
\end{document}